\documentclass{mem}
\input{psfig}
\usepackage{natbib}
\usepackage{txfonts}\usepackage{balance}
\usepackage{graphicx}
\usepackage[a4paper]{hyperref}
\idline{1}{1}
\begin{document}
\def\teff{$T\rm_{eff }$}
\def\kms{$\mathrm {km s}^{-1}$}

\title{
CoRoT and asteroseismology.
}

   \subtitle{Preparatory work and simultaneous ground--based monitoring}

\author{
E.\,Poretti\inst{1} \and
M.\,Rainer\inst{1} \and
K.\,Uytterhoeven\inst{1} \and
G.\,Cutispoto\inst{2} \and
E.\,Distefano\inst{2} \and
P.\,Romano\inst{2}
          }

  \offprints{E. Poretti}

\institute{
INAF -- Osservatorio Astronomico di Brera, Via E. Bianchi 46,
I-23807 Merate, Italy.\\
\email{ennio.poretti,monica.rainer,katrien.uytterhoeven@brera.inaf.it}
\and
INAF -- Osservatorio Astrofisico di Catania, Via S. Sofia 78,
I-95123 Catania, Italy.\\ \email {gcutispoto,eds,prom@oact.inaf.it}
}

\authorrunning{Poretti et al.}

\titlerunning{CoRoT ground--based observations}

\abstract{The successful launch of the CoRoT (COnvection, ROtation and planetary Transits) 
satellite opens a new era in asteroseismology. The space photometry is complemented
by high--resolution spectroscopy and multicolour photometry from ground, to disclose 
the pulsational content of the asteroseismic targets in the most complete way. 
Some preliminary results obtained with both types of data are presented. 
}
\maketitle{}

\section{Introduction}

The satellite CoRoT (COnvection, ROtation and planetary Transits) 
has been successfully launched from Ba\"{\i}konur into a nearly perfect orbit   
by a new Soyuz~II--1-b rocket on December 27, 2006. It performs
photometric observations of stars with unprecedented high accuracy.
Its goal is twofold: the study of stellar interiors (the asteroseismic part) and the
search for extrasolar planets (the exoplanetary part). 
The CoRoT mission is a French--led one (75\%; the Principal Investigator is
Annie Baglin, Meudon Observatory), but also 
Spain, Austria, Belgium, Germany, Brazil and  ESA are involved. 
After an initial participation which collected a wide interest in the national
community, the Italian Space Agency ASI withdrew the official support to the project.

The mission has an unique instrument: a 27--cm aperture telescope equipped with
two CCDs for each scientific case.
The selected CoRoT direction of pointing  is
a double--cone (the CoRoT eyes) centered at $\alpha$=6h50m/18h50m 
(galactic Anticenter/Center), $\delta=0^{\rm o}$;
the radius of each eye is 10$^{\rm o}$. 
To achieve its goals, CoRoT will uninterruptedly observe five
fields for 150~d each (long runs). The field--of--view  of each
pointing is 1.3x2.7~deg$^2$.
In the asteroseismic channel this will result in high--precision 
photometry, with an expected 
noise level in the frequency spectrum of 0.7~ppm (parts per million) over a 5~d 
time baseline for a $V\sim$6.0 star. 
\citet{eric} give a very detailed discussion of the Seismology Programme of CoRoT
and stress the importance of the asteroseismic approach to the open questions of 
stellar physics.

We remind that in the exoplanetary channel CoRoT monitor simultaneously up to 12000 stars
in each long run in the 11.0$<V<$16.0 range. Thus, during the expected 2.5--y satellite lifetime,
a total of 60,000 light curves will be produced at a sampling rate of 8~min. 
The extrasolar planets are detected by measuring the weak decrease
of the star flux due to the transit of a planet in front of the star disk. 
\citet{barge} describe the methods which will be used to extract the transit signature
from the data and also comment the scientific impact of the mission;
\citet{circeo} describe the Italian contribution to this exoplanetary research.

At the moment, all the on-board systems are working as efficiently as predicted,  in some
cases significantly better than expected. Therefore, it is not surprising that the analysis
of the raw data (i.e., not all the sources of noise have been taken into account yet and
removed from the data) already disclosed the first exoplanet, CoRoT-Exo-1b\footnote{see
http://www.cnes.fr/web/5891-corot-decouvre-sa-premiere-exoplanete-.php\label{cnes}}. 

\section{The survey of the fields around primary targets: the Serra La Nave contribution}

\begin{figure}
\centerline{\psfig{file=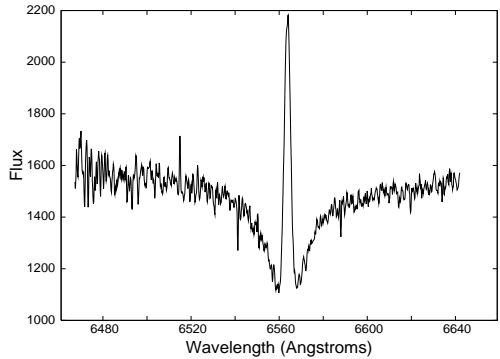,angle=270,width=6.7truecm}}
\caption{\footnotesize The H$\alpha$ line of HD~50087 shows the emission peak typical of
Be stars. The spectrum has been obtained at Serra La Nave Observatory with the FRESCO
spectrograph.
}
\label{sln}
\end{figure}

Despite the ASI withdrawal, a group of researchers still remained active on
various aspects of the CoRoT mission. \citet{canarie} described the Italian contribution
to the preparatory work to select the asteroseismologic targets and to build up
the GAUDI archive. In that part of the Italian contribution, the Telescopio
Nazionale Galileo played  a crucial role in completing the observational
tasks on time to respect the tight time schedule. After those observations, the short list
of the candidate primary targets was issued. We would like to stress here 
 a subsequent relevant contribution,
i.e., the observations carried out with the FRESCO (Fiber--optic Reosc Echelle
Spectrograph) instrument mounted at the 0.91--m Cassegrain telescope of the Serra
La Nave mountain station (1725~m above sea level) of the Osservatorio
Astrofisico di Catania. About 120 stars have been monitored in the spectral range 
from 430 to 680~nm ($R$=21,000 in the cross--dispersion mode).
Thanks to these observations
it has been possible to search for secondary targets located around the
already selected primary ones, particularly in the 8.0$<V<$9.5 range. Indeed,
this range was not covered by the previous surveys performed at ESO and
Haute Provence Observatory, where the limiting magnitude was $V<$8.0 in the complete area of the CoRoT eyes.
As an example of the usefulness of Serra La Nave observations, Figure~\ref{sln} shows 
the spectrum of a newly discovered Be star.

\section{The first asteroseismic targets}

Table~\ref{tarir}  lists the stars which have been observed in the asteroseismic CCDs
in the IR01, i.e., in the observing run (60~d long) immediately following the start of scientific
activities. 
The excellent CoRoT performances are proved by the detection
of the fingerprint of the asteroseismic oscillations, i.e., 
the regular spacing of the peaks in the power spectrum of a solar--like star$^{1}$.
As a preparatory work for the CoRoT mission, we planned spectroscopic 
observations to detect line--profile variations in the spectra of the 
asteroseismic targets.  We are applying to the CoRoT targets the know--how we acquired in several
years of ground--based studies on some classes of 
pulsating stars ($\delta$ Sct, $\beta$ Cep, $\gamma$ Dor, Be,~...). 
In particular, a Large Programme has been granted
at the FEROS@2.2m ESO-MPI instrument (fifteen nights per semester for 
four consecutive ESO periods), covering the first 1.5~y of the CoRoT lifetime. 
Moreover, other Large  Programmes will
be carried out at the  Observatoire de Haute Provence (using the new SOPHIE@1.92m instrument)
and at the Calar Alto Observatory (using the FOCES@2.2m instrument).

\begin{figure}
\centerline{\psfig{file=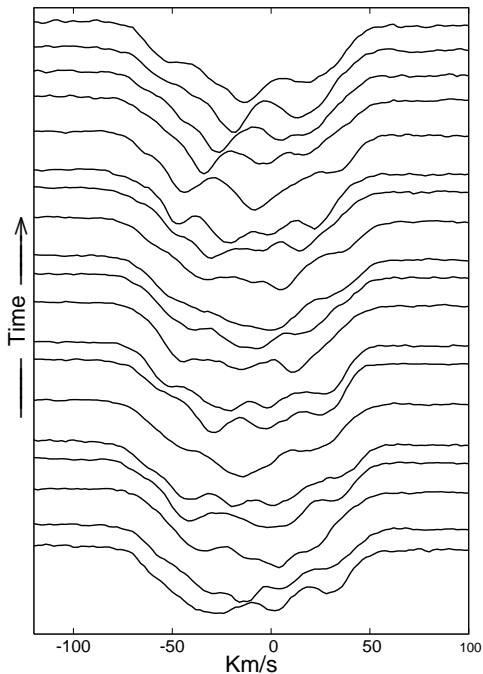,width=6.5truecm}}
\caption{\footnotesize The signature of nonradial modes are clearly visible
in the average profiles of the
$\delta$ Sct star HD~50844 on the night of January 6th, 2007. The observations were
carried out with the FEROS instrument mounted on the 2.2m ESO-MPI telescope at La Silla.
}
\label{delta}
\end{figure}

Only the  analysis of the line--profile variations will allow us to identify the
excited  modes, since in the stars located in the instability strip and on the
upper Main Sequence the regular 
spacing is destroyed by the mode selection operated by nonlinear effects. 
Figure~\ref{delta} shows an example of evident line--profile variations in the spectra
of the $\delta$ Sct star HD~50844. The mean line--profile 
has been calculated for each $R$=48,000~spectrum; the nonradial modes are seen
as propagating waves across the spectral line. For an example of mode identifications in
$\delta$ Sct stars see \citet{zima}.
\begin{table}
\caption{Targets of the IR01, ordered according to decreasing
brightness. The stars observed at ESO are in italic.}
\begin{tabular}{rr ll l}
\hline
\noalign{\smallskip}
\multicolumn{2}{c}{Star}&\multicolumn{1}{c}{$V$} &
\multicolumn{1}{c}{Sp.} & \multicolumn{1}{c}{Notes} \\
\noalign{\smallskip}
\hline
\noalign{\smallskip}
{\it HD}&{\it 50747} & 5.45 & A4 &  SB2 \\
HD & 49933 & 5.78  & F2V   &    \\
HD & 50890 & 6.03  & G6 &   Giant \\
HD & 50820 & 6.27 & B3IV   &   Be, SB2 \\
HD & 50170 & 6.86  & F2 &          \\
{\it HD}&{\it 51106} & 7.35 & A3m &  SB2 \\
{\it HD}&{\it 50846} & 8.43  & B5   & Eclip. variable \\
{\it HD}&{\it 50844} & 9.09  & A2 & $\delta$ Sct \\
HD & 50773 & 9.36  & A2 &        \\
{\it HD}&{\it 292790} & 9.48 & F8 &     \\
\noalign{\smallskip}
\hline
\hline
\end{tabular}
\label{tarir}
\end{table}

\begin{figure}
\centerline{\psfig{file=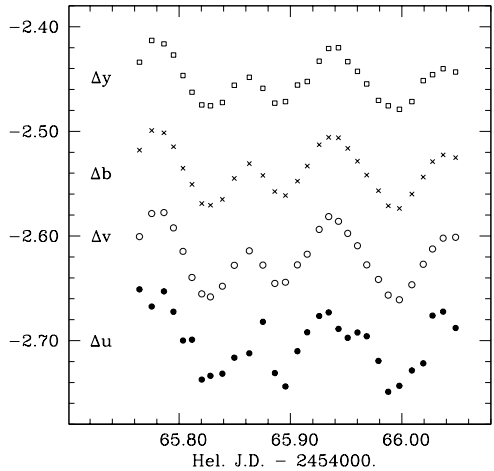,width=7.0truecm}}
\caption{\footnotesize The $uvby$ light curves of the $\delta$ Sct star HD~50844 obtained on the night of November 25, 2006
at S.~Pedro M\'artir Observatory. The magnitude differences have an arbitrary zeropoint.}
\label{spm}
\end{figure}
In Table~\ref{tarir} the targets observed during the first runs at ESO, OHP and Calar Alto
are indicated in italic. In addition to HD~50844, line--profile variations have been detected
in the spectra of the $\gamma$ Dor HD~49434 and of the Be star HD~50209. 
All the FEROS spectra have been reduced by the Brera team and delivered to the teams
responsible for the data analysis of the respective stars (Nice, Leuven, Granada, and
Meudon).  Also the observations
of the binary stars, carried out mainly to obtain a radial velocity curve and subsequently an orbital
solution, revealed intriguing features: HD~50747 is a triple system and one of the components
of HD~50846 is  a Be star.

In addition to high--resolution spectroscopy, we continued to monitor the asteroseismic
targets in multicolour photometry. Indeed, the CoRoT photometry does not supply any colour
information, since it is unfiltered. It is well known that phase shifts and amplitude
ratios can supply useful constraints to identify pulsation modes \citep{rafa}. 
Figure~\ref{spm} shows an example of $uvby$ light curves obtained on HD~50844 at S.~Pedro
M\'artir Observatory; the accuracy of a single point is about 0.002~mag in the $vby$ filters
and around 0.008~mag in the $u$ filter. This photomeric project is performed using the observational
nights granted to Brera Astronomical Observatory and is flanked by a similar one
at Sierra Nevada Observatory (Spain); the projects use two twin Danish photometers.
The photometric observations from ground  allow detection of  terms with amplitudes of about
1~mmag \citep{vienna}, while CoRoT will arrive to 2--3 orders of magnitudes smaller
\citep{eric}. 
Therefore, complementary ground--based campaigns can help to identify only the terms with 
the largest amplitudes. 
However,  by combining the photometric results  with the spectroscopic ones we could 
put some useful and tight constraints on the theoretical models. We stress that CoRoT will supply
frequencies free from aliasing effects (thanks to the 150~d continuous observations) and
therefore the identification of the exact values of the frequencies is not a severe problem. 

\section{Conclusions}
During all the preparatory activities for the CoRoT space mission
the contribution from the Italian community has been very valuable,
allowing first the monitoring of stars with $V<$8.0 in the CoRoT eyes
and then that  of fainter secondary targets located closely to the
primary ones. The use of national (Telescopio Nazionale Galileo) and
local (Serra La Nave Observatory) facilities has been decisive and
crucial.  Also considering the full characterisation of the exoplanetary
fields and the study of the interaction between stellar activity and planetary
transit detection \citep{circeo},
the Italian researchers provided original and useful inputs 
to the scientific profile to the mission.

New contributions and a very promising feedback in terms of understanding of
the stellar physics are expected from the full exploitation of the ground--based
monitoring of the asteroseismic targets.
\begin{acknowledgements}
The authors wish to thank once more all the Italian researchers and students 
(post--doc, PhD, undergraduate)
who are participating to the CoRoT activities; they  bear witness of the large interest of our
community into this exciting and pioneering space mission. 
K.~Uytterhoeven acknowledges the support of the European Community under the Marie Curie Intra-European
Fellowship, Contract 024476--PrepCOROT.
\end{acknowledgements}

\end{document}